\documentclass[fleqn,usenatbib]{mnras}

\usepackage{newtxtext,newtxmath}
\usepackage[T1]{fontenc}
\usepackage{ae,aecompl}

\usepackage{graphicx}	% Including figure files
\usepackage{amsmath}	% Advanced maths commands

\usepackage{amssymb}	% Extra maths symbols
\usepackage{bm}    
\usepackage{url}
\usepackage{subfig}
\usepackage{longtable}
\usepackage{booktabs,caption}
\usepackage[flushleft]{threeparttable}
\usepackage{fixltx2e}
\usepackage{adjustbox}
\usepackage{multirow}
\usepackage{ltablex}

\newcommand{\ToolName}{\texttt{lenstruction}}

\title[]{The evolution of the size-mass relation at $z$=1-3 derived from the complete Hubble Frontier Fields data set}

\author[L. Yang et al.]{
Lilan Yang,$^{1,2}$\thanks{E-mail: ylilan@astro.ucla.edu}
Guido Roberts-Borsani,$^{2}$
Tommaso Treu,$^{2}$
Simon Birrer,$^{3}$
Takahiro Morishita,$^{4}$
\and
Maru\v{s}a Brada\v{c}$^{5}$
\\
\\
\\
$^{1}$School of Physics and Technology, Wuhan University, Wuhan 430072, China \\
$^{2}$Department of Physics and Astronomy, University of California, Los Angeles, CA 90095-1547, USA\\
$^{3}$Kavli Institute for Particle Astrophysics and Cosmology and Department of Physics, Stanford University, Stanford, CA 94305, USA\\
$^{4}$Space Telescope Science Institute, 3700 San Martin Drive, Baltimore, MD 21218, USA\\
$^{5}$Department of Physics, University of California, Davis, CA 95616, USA\\
}

\date{Accepted XXX. Received YYY; in original form ZZZ}

\pubyear{2020}

\begin{document}
\label{firstpage}
\pagerange{\pageref{firstpage}--\pageref{lastpage}}
\maketitle

% Abstract of the paper
\begin{abstract}

We measure the size-mass relation  and its evolution between redshifts 1$<z<$3, 
using galaxies lensed by six foreground Hubble Frontier Fields clusters. 
The power afforded by strong gravitation lensing allows us to observe galaxies with higher angular resolution beyond current facilities. 
We select a stellar mass limited sample 
and divide them into star-forming or quiescent classes based on their rest-frame UVJ colors from the ASTRODEEP catalogs. 
Source reconstruction is carried out with the recently-released \texttt{lenstruction} software,
which is built on the multi-purpose gravitational lensing software \texttt{lenstronomy}.
We derive the empirical relation between size and mass for the late-type galaxies with $M_{*}>3\times10^{9}M_{\sun}$ at 1$<z<$2.5 and $M_{*}>5\times10^{9}M_{\sun}$ at 2.5$<z<$3, 
and at a fixed stellar mass, we find galaxy sizes evolve as $R\textsubscript{eff} \propto  (1+z)^{-1.05\pm0.37}$.
The intrinsic scatter is $<0.1$ dex at $z<1.5$ but increases to $\sim0.3$ dex at higher redshift. 
The results are in good agreement with those obtained in blank fields.
We evaluate the uncertainties associated with the choice of lens model by comparing size measurements using five different and publicly available models, 
finding the choice of lens model leads to a 3.7\% uncertainty of the median value, and $\sim 25$ \% scatter for individual galaxies.
Our work demonstrates the use of strong lensing magnification to boost resolution does not introduce significant uncertainties in this kind of work, and paves the way for wholesale applications of the sophisticated lens reconstruction technique to higher redshifts and larger samples.

\end{abstract}

% Select between one and six entries from the list of approved keywords.
% Don't make up new ones.

%%%%%%%%%%%%%%%%%%%%%%%%%%%%%%%%%%%%%%%%%%%%%%%%%%

%%%%%%%%%%%%%%%%% BODY OF PAPER %%%%%%%%%%%%%%%%%%

\section{Introduction}
A fundamental evolutionary parameter of a galaxy is its size and the size distribution of galaxy populations can reveal important clues as to their assembly histories and underlying dark matter halos \citep{Mo1998, Wechsler2018}. The comparison of galaxy structural properties such as sizes, stellar masses and luminosities further reveals tight scaling relations which likely dictate complex and diverse evolutionary pathways and afford the means to test the standard paradigm of galaxy formation and evolution. As an example, the size evolution of different galaxy types (i.e., early- and late-type galaxies) is found to be remarkably different, suggesting clearly distinct modes of stellar growth and dark matter halo assembly.

The empirical size-mass relationship, defined as $R\textsubscript{eff} \propto  M^{\alpha}$, and its evolution with redshift, $R\textsubscript{eff} \propto (1+z)^{\beta}$, 
have been investigated by several previous studies for both early- and late-type galaxies, sometimes with conflicting end results that can aid to test the basic theory of galaxy formation \citep{Mo1998}. For instance, setting the benchmark at ``zero redshift'' using a complete sample of $\sim$140,000 local galaxies (both early-type and late-type) from the Sloan Digital Sky Survey, \cite{Shen2003} found a power-law slope of $\alpha\sim$0.4 for late-type galaxies more massive than M$_{*}>10^{10.6}M_{\sun}$), while below this characteristic stellar mass the slope flattens down further to $\alpha\sim$0.15, implying a less rapid size evolution with stellar mass. On average, they find more massive galaxies tend to be characterised by larger radii than their less massive counterparts, implying a degree of ``inside out'' galaxy growth. Comparing this to their samples of early-type galaxies, the authors found that the relation for early-type galaxies displays a significantly steeper slope at fixed stellar mass, with $\alpha\sim$0.55, indicating a potentially separate and much faster evolutionary pathway. Despite the narrow redshift range of their sample (0.05$<z<$0.15), the authors conclude a negligible change in their size-mass relations over redshift. 

Further expanding the analysis to redshifts of 0$<z<$3 with $\sim$31,000 galaxies of M$_{*}\gtrsim$10$^{9}$M$_{\odot}$ from CANDELS and 3D-HST, \cite{vanderwel2014} found $\alpha$ values ($\alpha\sim$0.2) for both early- and late-type galaxies which are consistent with those reported for the low-redshift Universe, however the size distributions of these galaxies are found to be significantly smaller - i.e., a factor of $\sim$2 and $\sim$4 for late- and early-type galaxies respectively - indicating some redshift evolution ($\beta\sim-$0.75 and $\beta\sim-$1.48). These findings were reaffirmed by \cite{Morishita2014} and \cite{Mowla2019}, who extended the work of \cite{vanderwel2014} to samples of higher mass galaxies at the same redshifts.

Despite the success of the aforementioned studies in characterising the size-mass relation from the local to the high-redshift Universe, importantly they are limited by the angular resolution of the Hubble Space Telescope (HST).
To circumvent this issue, gravitational lensing serves as a powerful tool with which to extend the investigation of scaling relations to potentially fainter and smaller galaxies.
The Hubble Frontier Fields (HFF) program \citep{Coe2015, Lotz2017} has delivered significant samples of extremely faint galaxies, pushing down to unprecedented depths and out to very high redshift ($z\sim$6-8), allowing for more detailed characterisation of reionization-era objects. In particular, size measurements of the faintest such galaxies are crucial for understanding the contributions of galaxies to the reionization process of the universe: parameterisation of the faint-end ($M_{UV}>$-15) of the $z>6$ galaxy luminosity functions (LFs) allow one to determine the prevalence of the most abundant star-forming sources that are likely to be responsible for driving the reionization process \citep{Atek2015, Kawamata2015, Livermore2017}. While extremely small sizes have been found for the faintest such galaxies (e.g., $\sim$200 pc; \citealt{Bouwens2017}), the associated uncertainty of these lensed measurements are driven by the assumed lens model and serve as the main source of uncertainty in the determination of the LF faint-end slope
 \citep{Grazian2011, Oesch2015, Alavi2016, Bouwens2017, Atek2018}.
In addition to the detection of especially faint sources, gravitationally lensing also provides the chance to observe galaxies with high angular resolution.
Pioneering works e.g., \cite{Marshall2007} presented study of super-resolving galaxies at $z\sim$0.5.
\cite{Newton2011} explored 46 strongly lensed galaxies from the Sloan Lens ACS Survey (SLACS) at lower redshift 0.4$<z<$0.8, and found sizes of $\sim$300pc for the lowest mass galaxies with the size-mass relation offset to smaller sizes w.r.t. blank fields. A similar result was found by \citet{Oldham2017} who studied a population of red lensed galaxies, demonstrating that they fall below the size mass relation, suggesting that they are the evolved version of the compact massive galaxies found at high redshifts. 
Recently, \cite{Vanzella2020} discovered a strongly lensed (magnification factor $\sim$40) Ly$\alpha$ emission galaxy at $z\sim6.63$, with
intrinsic effective radius $<150$ pc. 
However, those studies were focused on strongly lensed galaxies with magnification factor $\sim$10 or even larger, and thus possibly affected by magnification bias, which favors more compact sources.

Here, we build on these previous studies by studying the galaxies lensed by the HFF clusters,
with the intrinsic stellar mass $M_{*}>10^{9}M_{\sun}$ at 1$<z<$3.
Respect to previous lensing works, we study a complete stellar mass selected sample with a range of magnification, 
thus providing a statistically equivalent counterpart to blank field studies, at higher effective angular resolution.

In the context of lensed galaxies, lens models are necessary to determine the intrinsic source properties.
In the current state-of-art of lensing modeling technique in the cluster scale, lens models tend to reproduce the position of multiple images rather than hand a full source reconstruction.
Several works (e.g.,\cite{Bouwens2017, Atek2018}) presented the impact of the choice of lensing models on scaling relation and LF,
although \cite{Meneghetti2017} have confirmed the accuracy and precision of the different strong lensing methods to a certain degree.
Therefore, we have to consider systematic uncertainties due to the selection of lens models when analysis lensed galaxies.

In this manuscript, we investigate the size-mass relation and size evolution 
of both early- and late-type galaxies with  super-resolving galaxies derived from the HFF at 1$<z<$3, making use of package \ToolName\ \footnote{\url{https://github.com/ylilan/lenstruction}} developed by \citet{Yang2020}, which is powered by \texttt{lenstronomy}\footnote{\label{lenstronomy1}\url{https://github.com/sibirrer/lenstronomy}}\citep{Birrer2015, Birrer&Amara2018},
a multi-purpose open-source gravitational lens modeling package.
In Section~\ref{sec:data}, we describe the data and selection criteria.
In Section~\ref{sec:size-determ}, we describe the details of measuring the sizes of galaxies.
Evolution of size-mass relation is presented in Section~\ref{sec:evolution}.
Assessment of strong lensing uncertainties and discussion are in Section~\ref{sec:discussion}.
Finally, we give our summary in Section~\ref{sec:summary}.
We assume the standard $\Lambda$CDM cosmology with parameters $(\Omega_{M}, \Omega_{\Lambda}, h)=(0.27, 0.73, 0.71)$, 
AB magnitude and the ~\cite{Chabrier2003} stellar initial mass function.

%section2
\section{Data}\label{sec:data}

\subsection{Multiwavelength photometric catalogues}\label{sec:astrodeep}
We base our analysis on the v1.0 reductions of HFF imaging data and the latest publicly-available lens models\footnote{\url{http://www.stsci.edu/hst/campaigns/frontier-fields/Lensing-Models}}. The HFF program provides ultra-deep observations over six lensing clusters, Abell 2744, MACS J0416.1-2403, MACS J0717.5+3745, MACS J1149.5+2223, Abell S1063 and Abell 370, and obtains images in HST ACS (F435W, F606W, F814W) and WFC3/IR (F105W, F125W, F140W, F160W) on both the main cluster and the parallel fields.
We adopt the multiwavelength photometric catalogues of these images from the ASTRODEEP project \footnote{\url{http://www.astrodeep.eu/}}, 
which collects imaging
from HST to ground-based $K$-band and Spitzer/IRAC, 
and provides photometrically-derived global galaxy properties (e.g., redshift, stellar mass, star formation rate) from several studies derived with SED-fitting codes \citep{Merlin2016, Castellano2016, Criscienzo2017, Shipley2018, Bradac2019}. 
Prior to their use, we correct the observed stellar masses for magnification using our fiducial lens model (see Section \ref{sec:lensing} below) 
and convert them to a \citet{Chabrier2003} IMF. 
Henceforth, we also refer to the F105W, F125W, F140W and F160W filters as $Y_{105}$, $J_{125}$, $JH_{140}$ and $H_{160}$, respectively.

%subsection2.1
\subsection{The HFF lens models}\label{sec:lensing}
Lens models for each of the six HFF central clusters have been provided to the community by five independent groups (\citealt{ Bradac2005, Williams2016Sebesta, Liesenborgs2007, Limousin2016, Zitrin2015, Sharon2014Johnson}; henceforth, Brada\v{c}, Williams, CATS, Zitrin and Sharon, respectively), each of which adopted their own techniques (e.g., parametric and non-parametric method) to derive the lens models. Most such models are exclusive to the main cluster, however some extend to the associated parallel field. We assume the Brada\v{c} model as our fiducial model and adopt its magnification to correct the stellar mass derived from ASTRODEEP. We note and will show that the choice of fiducial lens model introduces only very small differences and thus does not affect any of our conclusions.

The initial cluster-scale lens models are known as not accurate enough for elaborate source reconstruction,
 but having multiple lensed images, we can constrain the relative lensing operators via information of relative morphology contained in observed lensed images \citep{Yang2020}. 
In our sample, we have a small number of galaxies that are from spectroscopically-confirmed multiply-imaged systems. 
We consult the multiply-imaged systems with high-confidence from from \citet{Mahler2018}, \citet{Hoag2016}, \citet{Schmidt2014}, \citet{Treu2016}, \citet{Strait2018} and \citet{Caminha2016} over each of the six clusters, and there  are 8 multiply-imaged systems in our sample see Table~\ref{table:data}.

%subsection2.2
\subsection{Sample Selection}\label{sec:selection}
We obtain our sample of galaxies from the ASTRODEEP catalogs at redshifts 1$<z<$3 by selecting objects satisfying the following criteria:

\begin{itemize}
\item has a reliable photometric redshift classified by RELFLAG=1 in the ASTRODEEP catalogs,
\item has an intrinsic stellar mass $M_{*}>10^{9} M_{\sun}$,
\item has coverage by each of the central lens models derived by the five teams described above,
\item does not reside close to a particularly bright object or at the edge of the instrument detector,
\end{itemize}

\begin{figure*}
\includegraphics[width=2\columnwidth]{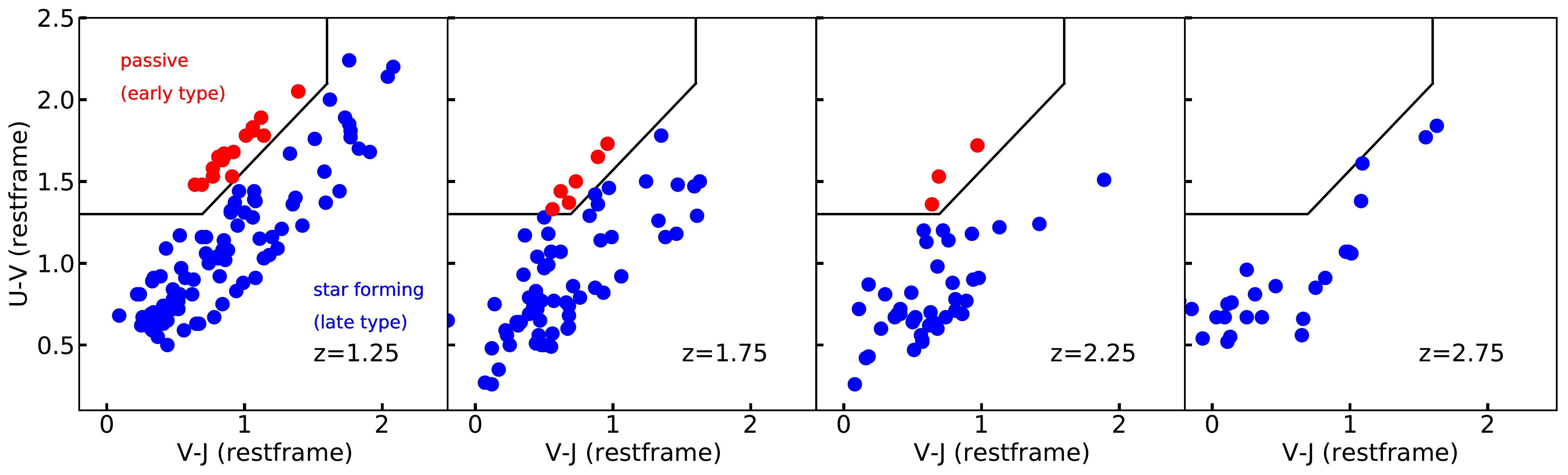}
\caption{The rest-frame UVJ color diagram for four redshift bins (each $\Delta z$=0.5 wide). 
Galaxies are classified into two types, passive (early-) or star-forming (late-type). 
The solid black lines in each panel indicates the selection criteria from \citep{Williams2009}, which is used in this work.}
\label{img:uvj}
\end{figure*}

\begin{figure*}
\includegraphics[width=2\columnwidth]{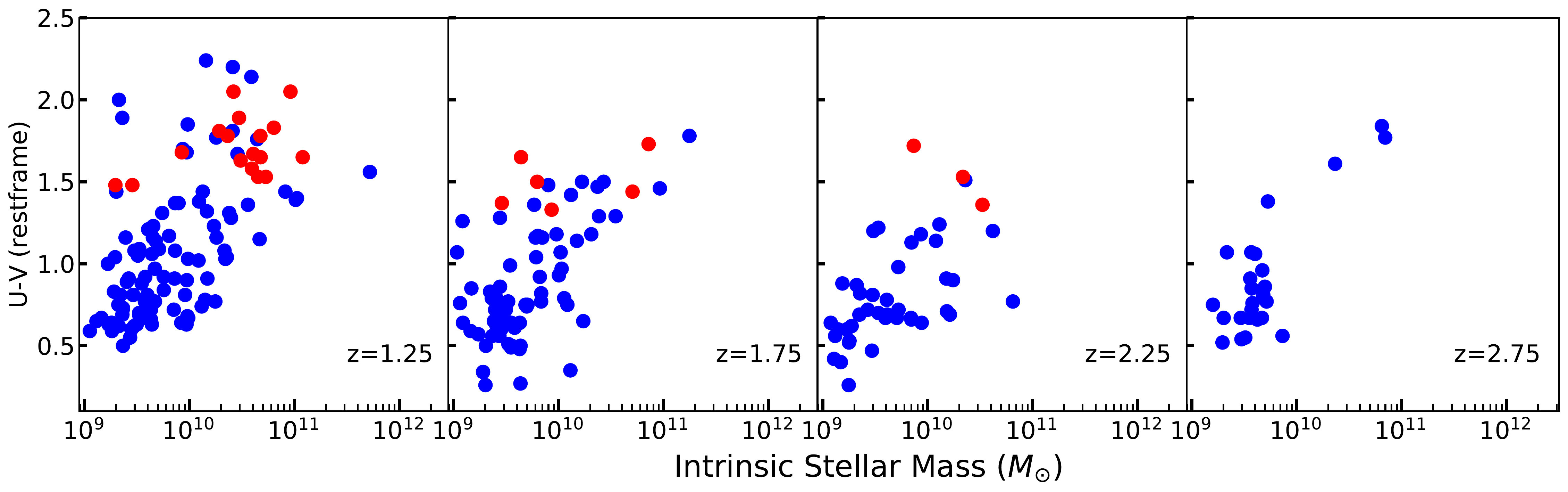}
\caption{The rest-frame U-V color as function of stellar mass for the same four redshift bins shown in Figure \ref{img:uvj}. Both figures adopt the same color scheme.}
\label{img:uvmass}
\end{figure*}

Applying the above set of criteria, we obtain a sample of 258 lensed galaxies at the aforementioned redshift, of which several objects have multiply-lensed images. 
In the later case, we count such galaxies only once, providing a final sample of 255 sources. 
We limit ourselves at this redshift range and leave the more complicated issue correlated with completeness at higher redshift to our future work.
Following the criteria presented by \cite{Williams2009}, we utilize rest-frame UVJ colors to distinguish between early-type (passive) and late-type (star-forming) galaxies, which we separate and present as a function of several redshift bins in Figure~\ref{img:uvj}. 
This color space allows for a separation of the two galaxy types, although we note beyond a redshift of 1.5 there are only a handful of early-type galaxies. In Figure~\ref{img:uvmass}, we present the same samples as a function of rest-frame U-V color and magnification-corrected stellar mass, where it becomes clear that passive galaxies tend to be redder at all redshift bins compared to their star-forming counterparts.

%figure1

%figure2

%table1

\begin{table*}
\centering
\caption{Catalog of lensed galaxies at redshift 1$<z<$3 which meet the selection criteria described in Section\ref{sec:selection}. 
The columns ID, coordinates, photometric redshift ($z\textsubscript{ph}$) and intrinsic stellar mass are shown from ASTRODEEP catalogs. The column $R\textsubscript{eff}$ represents the effective radius from MCMC analyses with uncertainties taken as the 16th and 84th percentiles, and
the column $n\textsubscript{s\'ersic}$ represents the S\'ersic index. The columns $z\textsubscript{spec}$,  multi-ID are the spectroscopic redshift and multiply-imaged system derived from references recorded in the last column.} \label{table:data} 
\resizebox{0.8 \textwidth}{!}{\begin{threeparttable}
\begin{tabular}{lccccccccc} 
\hline
\hline
ID & RA & Dec & $z\textsubscript{ph}$ & $z\textsubscript{spec}$ & M$_{*}$  & $R\textsubscript{eff}$  & $n\textsubscript{s\'ersic}$ & multi-ID & Ref $^a$\\
 &  &  &  & & ($\times$$10^{9}$M$_{\sun}$) & (kpc) &  &  &  \\
\hline
\multicolumn{10}{c}{cluster Abell 2744}\\
\hline

75 & 3.5893 & -30.4159 & 2.730 & ...& 3.79 & 0.79 $ ^{+0.00} _{ -0.00} $ & 1.39 & ... & ... \\
147 & 3.5770 & -30.4134 & 1.240 & ...& 12.19 & 2.27 $ ^{+0.00} _{ -0.00} $ & 1.00 & ... & ... \\
161 & 3.5753 & -30.4128 & 1.390 & ...& 1.69 & 6.55 $ ^{+0.07} _{ -0.06} $ & 1.62 & ... & ... \\
163 & 3.5998 & -30.4139 & 1.680 & ...& 3.32 & 3.76 $ ^{+0.01} _{ -0.02} $ & 4.00 & ... & ... \\
273 & 3.5746 & -30.4123 & 1.372 & ...& 2.30 & 2.43 $ ^{+0.01} _{ -0.02} $ & 3.50 & ... & ... \\
370 & 3.5789 & -30.4100 & 1.274 & ...& 1.45 & 1.40 $ ^{+0.01} _{ -0.01} $ & 1.00 & ... & ... \\
438 & 3.5708 & -30.4102 & 2.540 & ...& 2.92 & 1.20 $ ^{+0.00} _{ -0.00} $ & 1.03 & ... & ... \\
442 & 3.5767 & -30.4102 & 1.664 & ...& 2.03 & 1.71 $ ^{+0.01} _{ -0.01} $ & 1.42 & ... & ... \\
451 & 3.5862 & -30.4100 & 1.498 & 1.688 & 3.29 & 7.08 $ ^{+0.21} _{ -0.20} $ & 1.93 & 1.3 & M18 \\
501 & 3.6062 & -30.4085 & 1.355 & ...& 26.23 & 1.81 $ ^{+0.01} _{ -0.01} $ & 2.52 & ... & ... \\
\multicolumn{10}{c}{...}\\
\hline
\end{tabular}

\begin{tablenotes}
\small
\item $^{a}$ 
References M18, H16, S14, T16, S18 and C16 represent 
$gold$ category in \cite{Mahler2018} (see their Table A1),		
$gold$ category in \cite{Hoag2016} (see their Table 2),
GLASS project in \cite{Schmidt2014} (see their Table 1),
$gold$ category in \cite{Treu2016} (see their Table 3),
\cite{Strait2018} (see their Table 2), 
and \cite{Caminha2016} (see their Table 1), respectively.
(This table is available in the online journal. A portion is shown here for guidance regarding its form and content.)
\end{tablenotes}

\end{threeparttable}}

\end{table*}

%section3
\section{Intrinsic size determination}\label{sec:size-determ}
In order to measure the size of the lensed galaxies, we make use of the recently published Python package \ToolName\ \citep{Yang2020},
which is built on the strong lensing software \texttt{lenstronomy} \citep{Birrer2015, Birrer&Amara2018}.
\texttt{lenstronomy} provides the core functionalities of the modeling, including the reconstruction of the source light profile while simultaneously considering both lensing and blurring effects, and the Bayesian inference formalism. \texttt{lenstronomy} is the work horse underneath through which those tasks are executed.
\ToolName\ is the wrapper around \texttt{lenstronomy} that provides the interface to the specific cluster data products and sets up the specific tasks required to achieve reliable source reconstructions and lens model corrections in the cluster lensing regime.
We refer the reader to the GitHub repository for more general information about \ToolName\ and \texttt{lenstronomy}.
Throughout this section we first describe the suitability of \ToolName\ for source reconstruction, describe our modeling procedures and the handling of multiply-lensed systems in Section~\ref{sec:modeling},
and finally provide a comparison with previous results found with the popular \texttt{GALFIT} tool in Section~\ref{sec:galfit}.

%subsection3.1
\subsection{Modelling procedure}\label{sec:modeling}
Photometric data capture only a portion of the underlying spectrum of a galaxy and thus stellar and gaseous content. Thus, in order for a consistent comparison across our redshift bins, we apply our size measurements over the $Y_{105}$ image for sources at redshifts 1$<z<$1.5, over the $J_{125}$ image for 1.5$<z<$2,  over the $JH_{140}$ image for 2$<z<$2.5, and over the $H_{160}$ image for 2.5$<z<$3. 
For measurements of the point-spread function (PSF), we select a stellar object in {the field of the cluster.} 
In order to measure the reconstructed 2D light profile of the source in each image, we parameterise the profile as an elliptical S\'ersic profile. 
The free parameters of an elliptical S\'ersic profile include a surface brightness amplitude, effective radius (or half-light radius; $R\textsubscript{eff}$), S\'ersic index (n$_{\text{s\'ersic}}$), axis ratio (q), position angle, and the central position of the source. 
We assume a S\'ersic index range between 1 to 4 and an axis ratio between 0.1 and 1. 
While the light profile shape changes dramatically as the n$_{\text{s\'ersic}}$ changes from 1 to 4, there is no such obvious change in the light profile shape from 4 to higher values \citep{Graham2013}.

Most of our sources are singly-imaged galaxies, and we fix the initial lens model due to the degeneracy between the image and the lens model.
For a given lens model, we simulate each of the lensed images and convolve with the PSF in the image place. 
By comparing the simulated image with the observation and estimating the noise in each pixel in the image plane as a combination of a Gaussian background rms, $\sigma_{bkg}$ and the Poisson term scaled with the exposure time,
we are able to derive the best-fit free parameters of the applied S\'ersic profile.
\ToolName\ can automatically estimate and remove the background flux level making use of \texttt{Background2D} in package \texttt{photutils}.
We note the potentially significant impact of intra-cluster light (ICL) on the size estimation due to the complexity of obtaining an ICL-subtracted image (see details in \cite{Merlin2016, Castellano2016, Criscienzo2017, Shipley2018, Bradac2019}). 
To test the validity of the routine used in  \ToolName, we compare the effective radius obtained from ICL-subtracted images and the raw images, and find an excellent agreement within the errors.

\begin{figure}
 \centering
\subfloat[From left to right, we show the observed lensed images, the modeled lensed images, and the normalized residuals (i.e., divided by uncertainty)]{\includegraphics[width=\columnwidth]{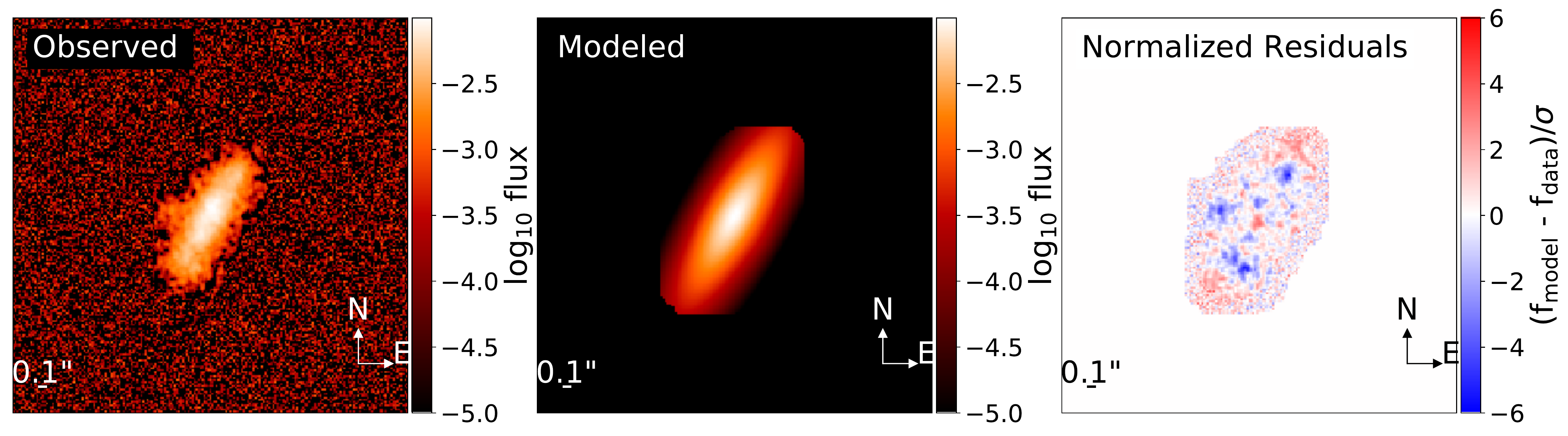} }\\
 \subfloat[Reconstructed source surface brightness distribution with lens models from Brada$\v c$.]
 { \includegraphics[width=0.55\columnwidth]{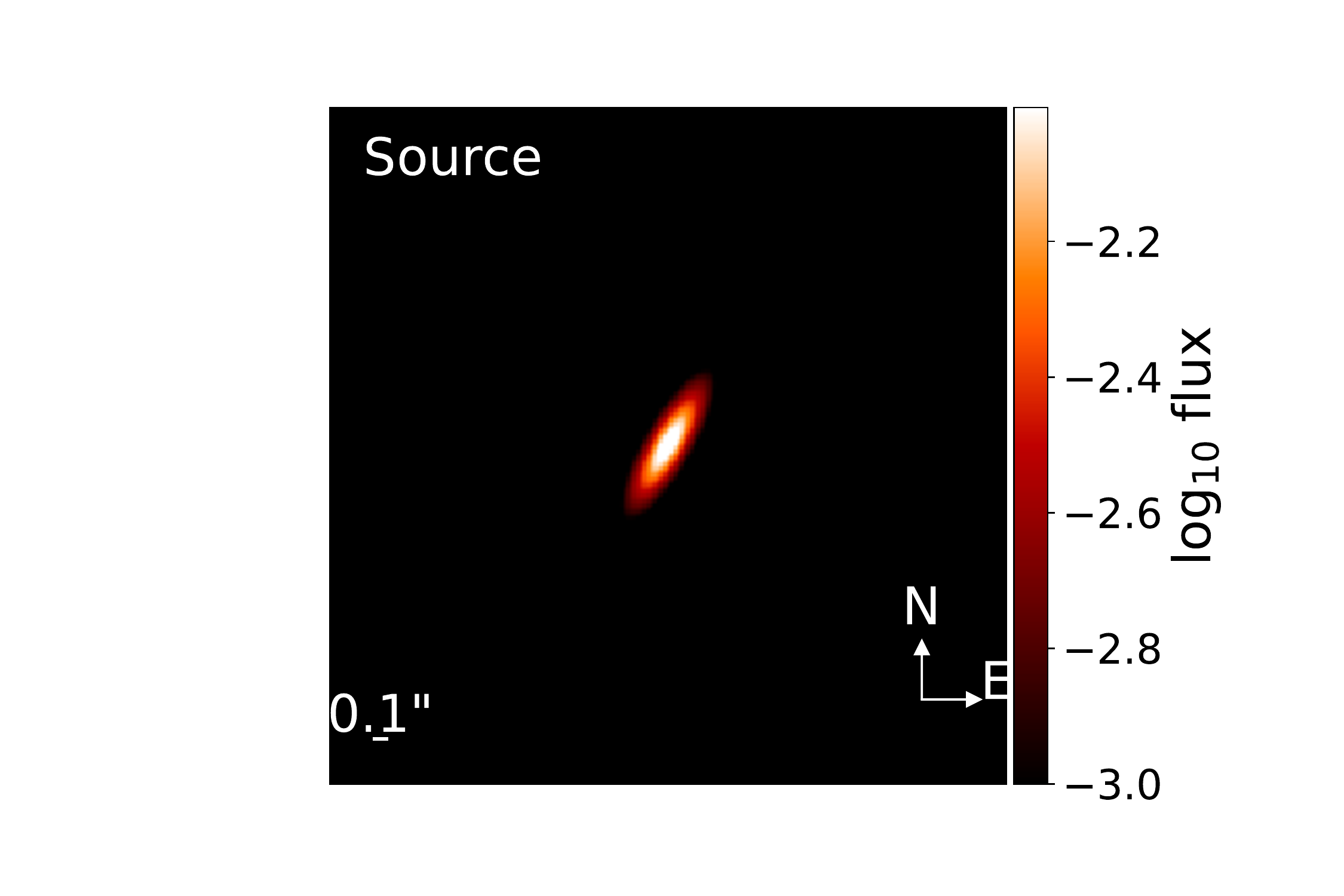} } 
\caption{Demonstration of the modeling results of the singly-imaged system using \ToolName /\textsc{lenstronomy}.}
\label{img:lenstruction}
\end{figure}

We present an example of modeling procedures in Figure ~\ref{img:lenstruction}.
The uncertainties on $R\textsubscript{eff}$ are taken as 16th and 84th values of the MCMC results. The best fits are visually inspected and a small number of fits are discarded based on catastrophic failures induced by nearby bright emission. We summarize the results of our fitting in Table~\ref{table:data}.

As mentioned in Section~\ref{sec:lensing}, there are 8 multiply-imaged systems in our sample (see Table~\ref{table:data}). When multiple images are available for a given source, one can obtain important constraints on the relative lensing operators. As such, in cases of a multiply-imaged galaxy, we fist correct the initial lens model before applying it: we fix the lens parameters of the least magnified image while allowing them to vary over the other images of the same object. In some cases, it is necessary to adopt a light profile of the source with high complexity, namely a shapelet with order up to 10 or even higher (for full details see \cite{Yang2020}). Finally, equipped with the corrected lens model, we perform the same modeling procedure as described above to obtain the size of the source.
The multiply-imaged systems offer an additional opportunity to quantify the uncertainty of the lens model itself, as discussed in our previous paper \citep{Yang2020}.

%subsection3.2
\subsection{Comparing size measurements:  \ToolName/\texttt{lenstronomy} vs \texttt{GALFIT} }\label{sec:galfit}
\begin{figure}
\includegraphics[width=\columnwidth]{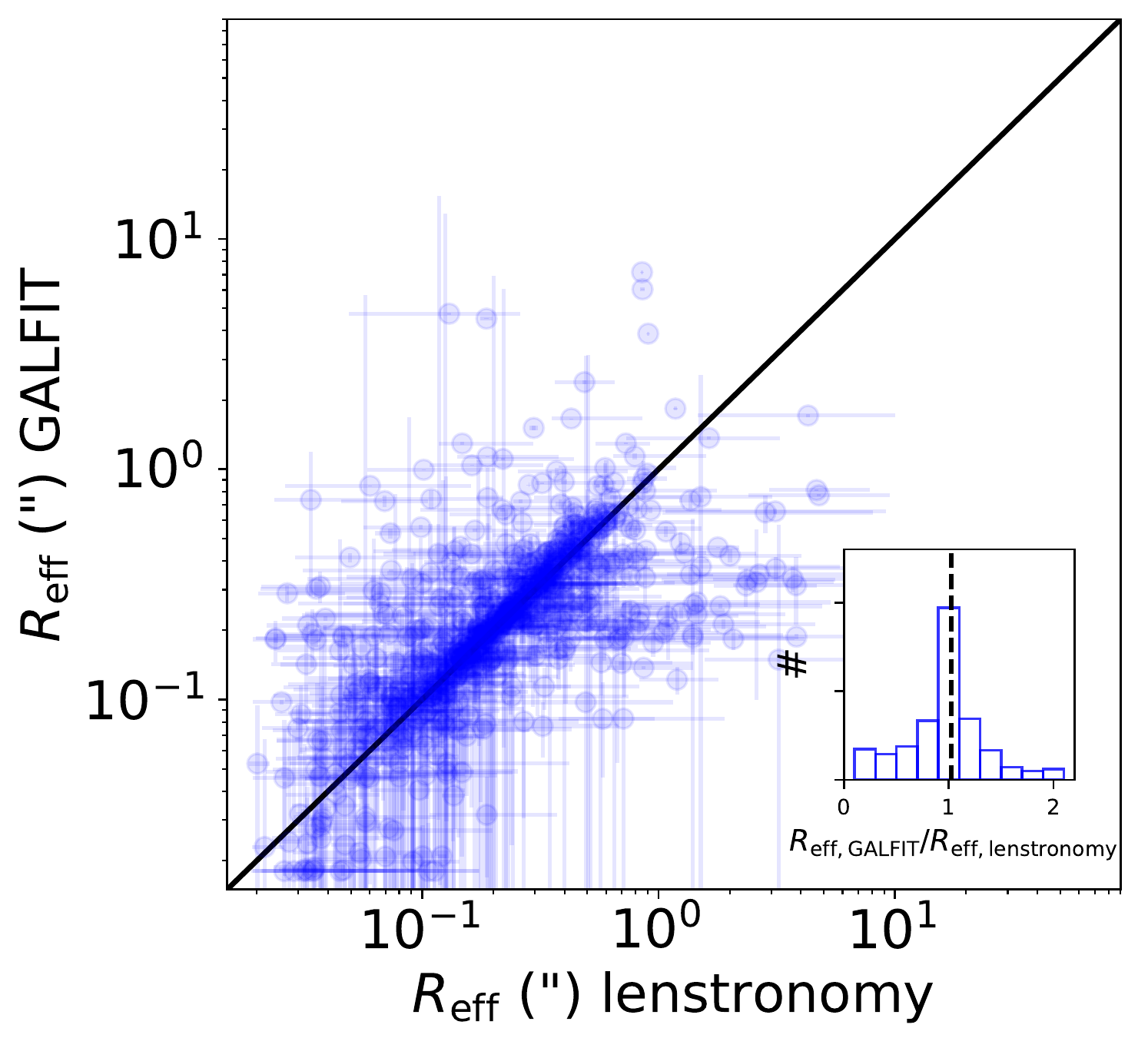}
\caption[]{Comparison of the size measured by the software \texttt{GALFIT} used by \cite{vanderWel2012} and by \texttt{lenstronomy} interfaced by \ToolName over a sample of $\sim$1000 randomly-selected unlensed galaxies from the CANDELS UDS field, using $H_{160}$-band images. The inner panel presents the distribution of size ratios between the two derivations, with a median ratio of 1.02$\pm$0.10. 
}
\label{img:uds-comp}
\end{figure}

As a consistency and reliability test of our measured galaxy sizes with performed by \texttt{lenstronomy} through \ToolName, we compare the size measurements between our code and \texttt{GALFIT}
over a test sample of $\sim$1000 randomly-selected field galaxies at same redshift range $1<z<3$, 
where the sizes for each of these galaxies was derived and presented by \citet{vanderWel2012} using the $H_{160}$ image (see their Table 2). 
To make a faithful comparison, we adopt a similar set of parameters as \citet{vanderWel2012}, e.g., S\'ersic index between 0.2 and 8, and axis ratio between 0.0001 and 1. 
We thus re-derive the sizes of the sample using the $H_{160}$ band and the same light profile model, and compare the results in Figure~\ref{img:uds-comp}. 
We find the size derivations between the two software yields very similar results, with a median size ratio of 1.02$\pm$0.10 for two samples (see the inner panel of Figure~\ref{img:uds-comp}). We conclude that our size derivations are thus robust and that such a comparison validates the use of \ToolName/\texttt{lenstronomy}.

%figure3

%figure4

%section4
\section{Evolution of size-mass distribution}\label{sec:evolution}
We show the intrinsic size-mass distribution of resolved, lensed galaxies from the FF as a function of redshift in Figure~\ref{img:bradac-sm}. Intuitively, we see that the well-known size-mass relation exists at all redshift bins for both early- (red color) and late-type galaxies (blue color) with large intrinsic scatter. Additionally, the early-type galaxies are on average smaller than the late-type galaxies. We perform a careful analytical description of the size-mass distribution in Section~\ref{sec:analytical} and provide an analysis of the size evolution in Section~\ref{sec:size-evo}.

%subsection4.1
\subsection{Analytical Description}\label{sec:analytical}
Following works by \citet{Shen2003}, \citet{vanderwel2014} and \citet{Mowla2019}, we assume the effective radius obeys a log-normal distribution and parameterize the size-mass scaling relation as,

%equation
\begin{equation} \label{re}
log\frac{R\textsubscript{eff} }{kpc}= log(A) + \alpha log\frac{M_{*}}{5\times10^{10}M_{\sun}}+N(\sigma),
\end{equation}
where A is the intercept, $\alpha$ is the slope, and $N(\sigma)$ is intrinsic scatter.

\begin{figure*}
\includegraphics[width=2\columnwidth]{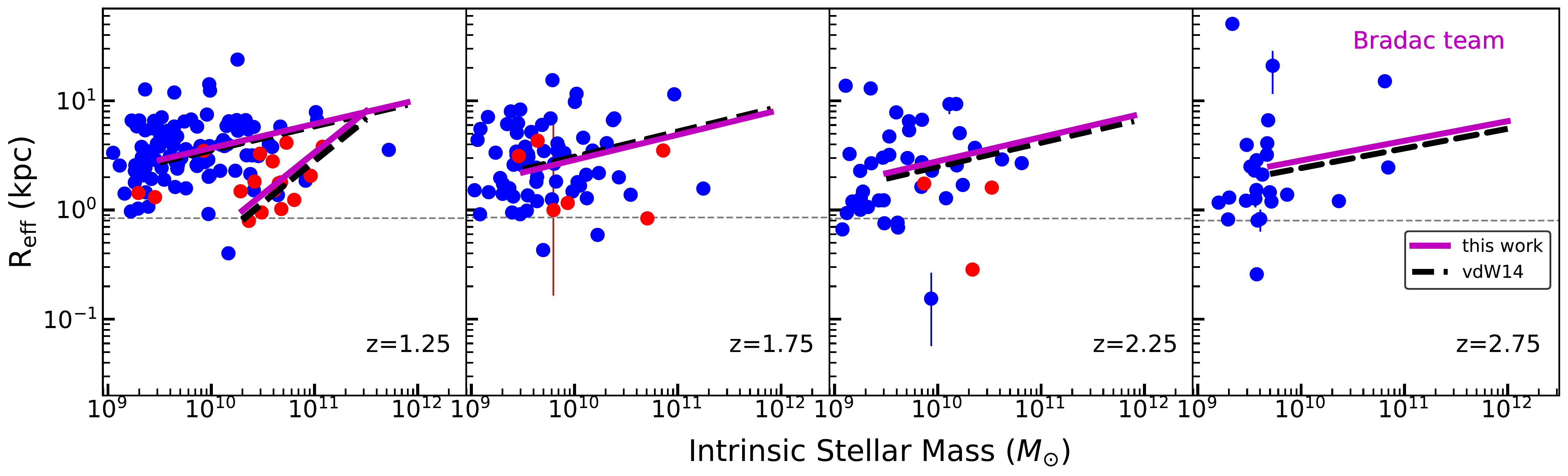}

\caption[]{Size-mass distribution of early- (red) and late-type (blue) galaxies at redshifts 1$<z<$3 assuming the fiducial lens model from the Brada\v{c} team.
The magenta lines indicate the fits to the data points (see also Table~\ref{table:size-mass}) while the black dash lines show the fitting results from \cite{vanderwel2014}.
The grey dash lines demonstrate the angular resolution limitation of the WFC3/IR.
}
\label{img:bradac-sm}
\end{figure*}

Given the small numbers of our sample of early-type galaxies, 
we only fit the galaxies at redshift $1<z<1.5$, and intrinsic stellar mass $M_{*} > 2\times10^{10}$, so that we avoid the potentially flatter part of the size-mass distribution at lower stellar mass.
For the late-type galaxies, where our sample size is larger, we fit galaxies with intrinsic stellar mass $M_{*} > 3\times10^{9}$ at $z<2.5$, and $M_{*} > 5\times10^{9}$ at $2.5<z<3$.
We believe that our mass-limited samples are likely to be complete.
Since several studies have found a lack of significant evolution of the slope, $\alpha$, with $\alpha\sim$0.7 for early-type and $\alpha\sim$0.2 for late-type galaxies \citep{Newman2012, vanderwel2014}, we opt to keep this parameter fixed at each redshift bin, as done in \citet{vanderwel2014}. 
With this assumption, we subsequently use a standard Bayesian approach to derive the posterior distribution of the parameters log(A) and $N(\sigma)$, while assigning 0.3 dex as the typical error in stellar mass throughout the fitting. The fitting results for both galaxy types are shown in Figure~\ref{img:bradac-sm} as magenta solid lines, adopting the fiducial lens model of the Brada\v{c} team, while we also include a black dashed line to represent the results of \citet{vanderwel2014}. For comparison, the fitting results are repeated for each of the five lens models used in this study and the resulting parameters are presented in Table~\ref{table:size-mass}. 
In each panel, the grey dash lines demonstrate the angular resolution limitation of WFC3/IR. 

%subsection4.2
\subsection{The redshift evolution of galaxy sizes and intrinsic scatter}\label{sec:size-evo}
\begin{figure*}
\includegraphics[width=\columnwidth]{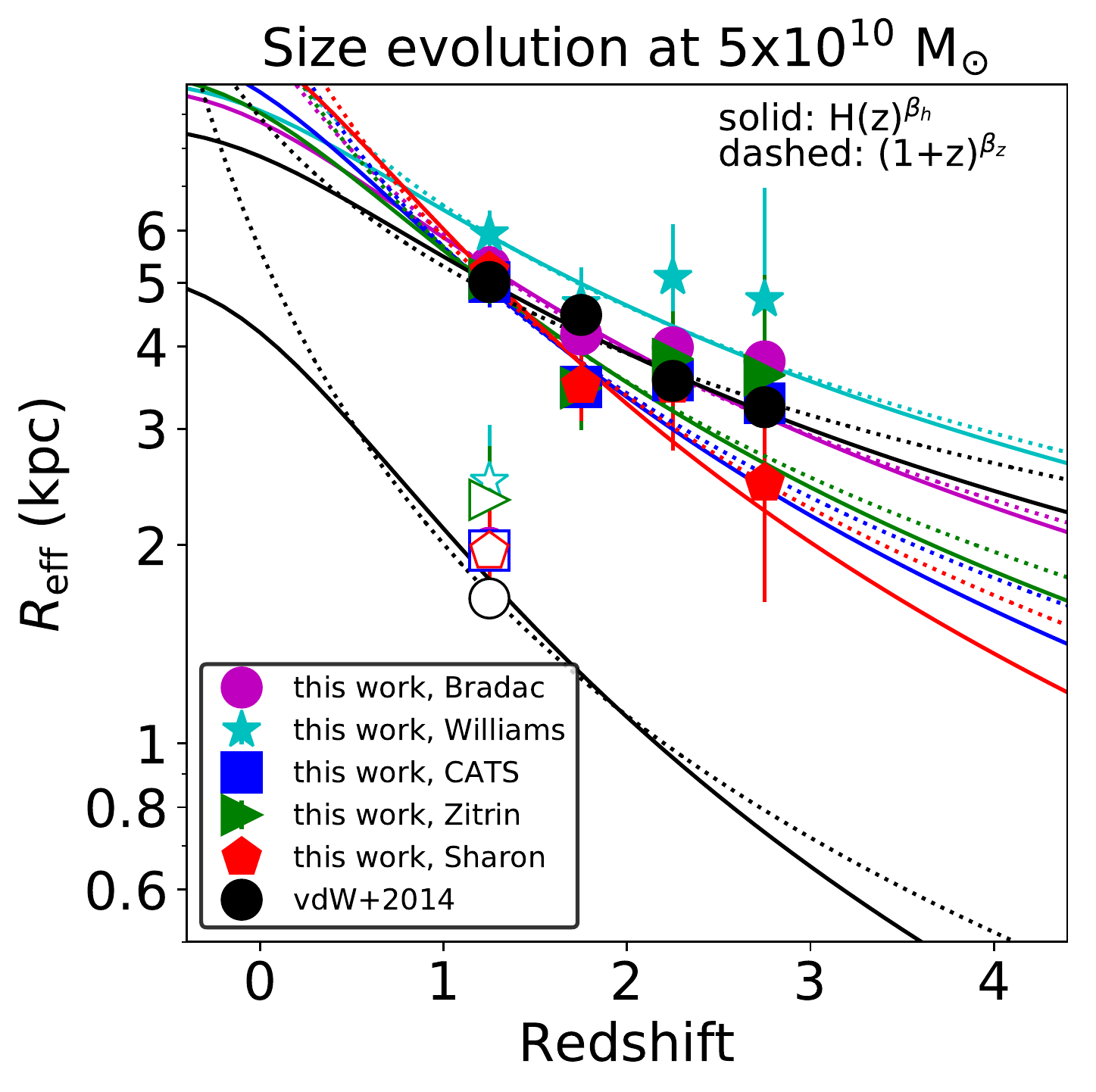}
\includegraphics[width=1.05\columnwidth]{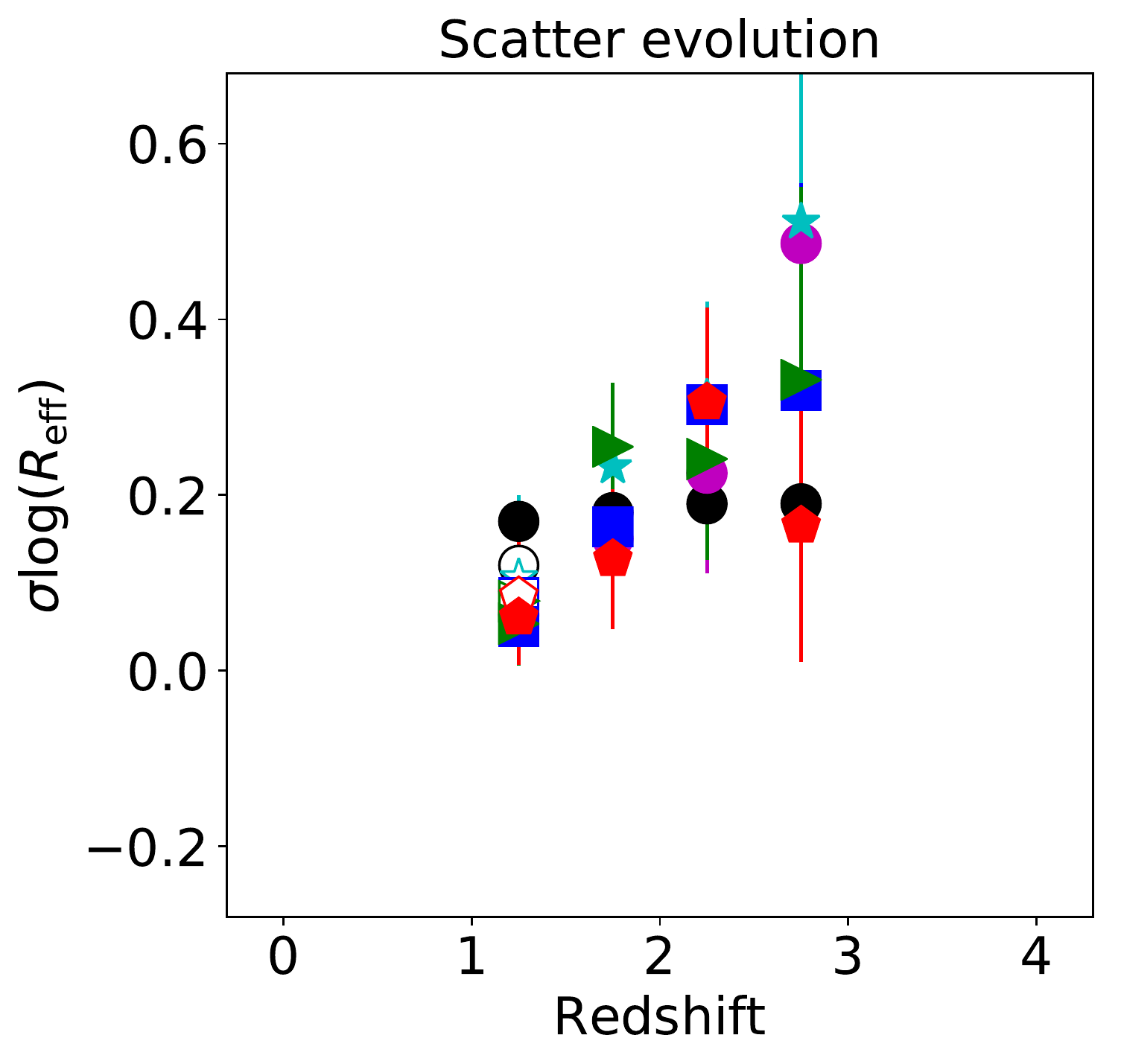}
\caption[]{Redshift evolution of galaxy sizes (left) and intrinsic scatter (right) and the former's parameterizations for each of the lens models used in this work. The filled and open symbols represent the results of the late-type and early-type galaxies, respectively, while the different colored lines represent the fitting results for different lens models while black lines show the results of \cite{vanderwel2014}, for comparison. Strong evolution is seen for the sizes of late-type galaxies and we parameterize such evolution as a function of $H(z)$ and $(1+z)$ shown by solid and dashed lines, respectively. 
For the intrinsic scatter, our uncertainty is too large to conclude whether there is evolution as a function of redshift.
}
\label{img:size_evo}
\end{figure*}

\begin{figure}
\includegraphics[width=0.8\columnwidth]{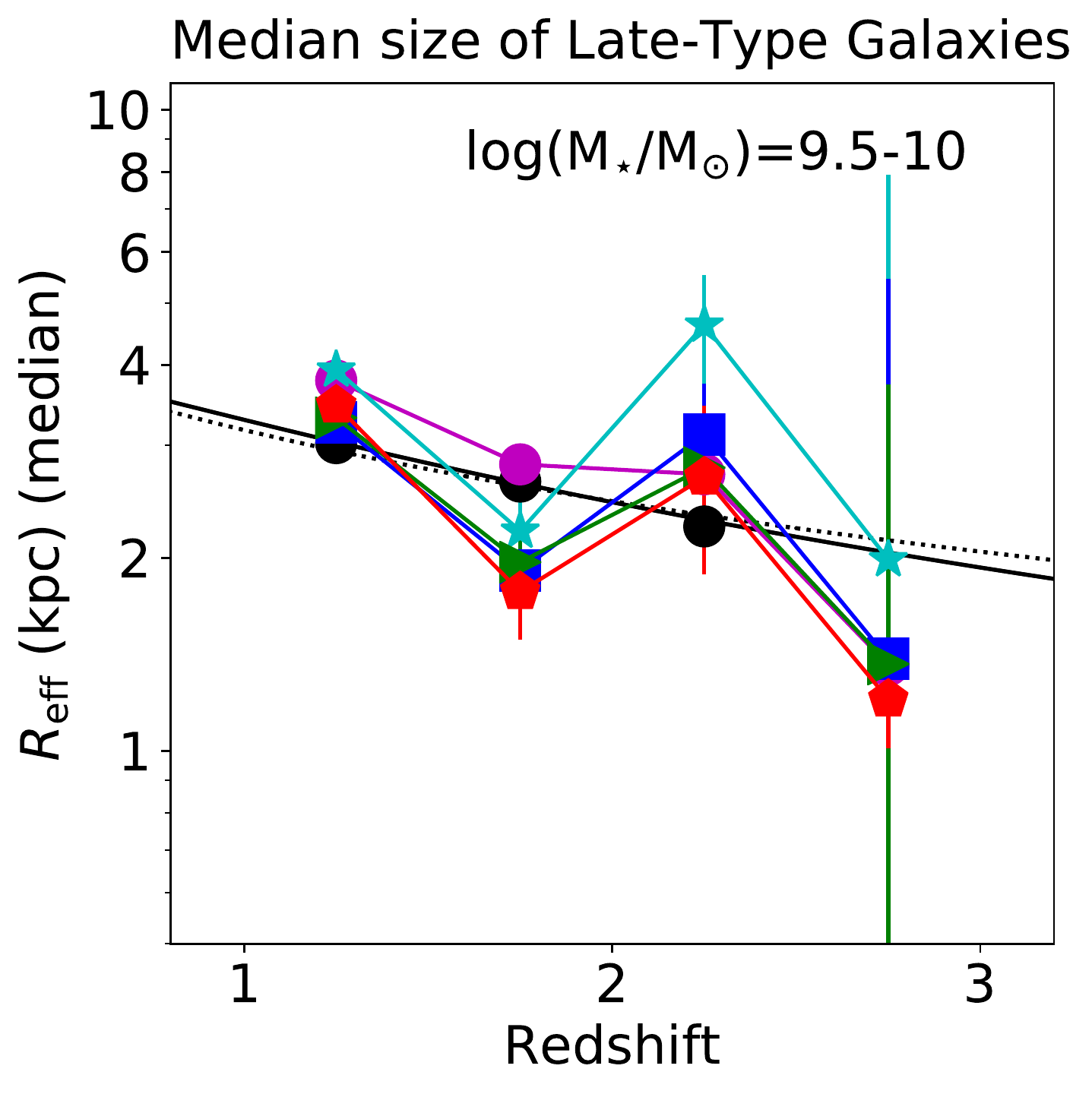}
\caption{Evolution of the median size of late-type galaxies with stellar mass log(M$_{*}$/M$_{\sun}$)=9.5-10. The color scheme is the same as in Figure~\ref{img:size_evo}.}
\label{img:median-size}
\end{figure}

The evolution of our measured galaxy sizes and their associated scatter as a function of redshift is presented in Figure~\ref{img:size_evo}. In the left panel, the intercept indicates the size evolution at a fixed stellar mass of $5\times10^{10}M_{\sun}$, with filled and open symbols representing late-type and early-type galaxies, respectively. Empirically, we parameterize the evolution of the y-axis intercept as a function of the cosmological scale factor, $\propto (1+z)^{\beta_z}$. From a more physical standpoint, since the size of a galaxy correlates with the underlying dark matter halo, we also parameterize it as function of the Hubble parameter $\propto H(z)^{\beta_h}$. The results of these parameterizations yield ${\beta_z}$=-1.05$\pm$0.37 and ${\beta_h}$=-0.80$\pm$0.37 for our fiducial lens model. We show these results and those for the other lens models as colored solid and dash lines, while summarizing them in Table~\ref{table:size-evo}. As in previous figures, for reference we show the results of field analyses from \cite{vanderwel2014} as black lines. 
In comparison, considering the larger uncertainties which due to the limitation of sample size, our results are consistent with \cite{vanderwel2014} for both early- and late-type galaxies.
We discuss the uncertainties of the ${\beta_z}$ and ${\beta_h}$ parameterizations as a function of lens model in Section~\ref{sec:lensmodel}. So far, we have presented the size evolution for a fixed stellar mass. For a range of stellar mass log(M$_{*}$/M$_{\sun}$)=9.5-10, we show the evolution of the median resulting size in Figure \ref{img:median-size}, again as a function of lens model. The trend in this figure shows similar behavior as the trends shown in the left panel of Figure~\ref{img:size_evo} and in each case is also consistent with the results from \cite{vanderwel2014}.

A possible evolution of the intrinsic scatter of late-type galaxies is investigated in the right panel of Figure~\ref{img:size_evo}. We notice a potential trend with the intrinsic scatter spanning only $<$0.1 dex at 1<$z$<1.5, increasing to $\sim$0.24 dex at 1.5<$z$<2.5, and finally reaching $>0.3$ dex for the majority of the lens models at 2.5$<z<$3. However it is not significant given our uncertainties.  
For reference, \cite{vanderwel2014} found little or no evolution for either late- or early-type galaxies, with a constant scatter of 0.16-0.19 dex and 0.1-0.15 dex respectively. A larger sample of galaxies is required to detect a trend, if present.

%table2
\begin{table*}
	\centering
	\caption[]{The fitting results of the size-mass distribution over all lens models, as shown in Figure~\ref{img:bradac-sm}. The slope $\alpha$ is fixed in the same fashion as \citep{vanderwel2014} (see their Table 1 for comparison).}
		\begin{tabular}{lcccccccc} 
		\hline
		\hline
		 &   & Late-type & &  & &Early-type &  &  \\\cline{2-4}\cline{6-8}			
 $z$ & log(A)  &$\alpha$ & $\sigma$ log($R\textsubscript{eff}$) & & log(A) &$\alpha$ &  $\sigma$ log($R\textsubscript{eff}$) & Lens model\\		
\hline		
 1.25 & 0.72 $\pm$ 0.04 & 0.22 & 0.06 $\pm$ 0.05& & 0.29 $\pm$ 0.09 & 0.76 & 0.09 $\pm$ 0.08 & Brada\v{c} \\
 1.75 & 0.62 $\pm$ 0.05 & 0.23 & 0.14 $\pm$ 0.08\\
 2.25 & 0.60 $\pm$ 0.08 & 0.22 & 0.21 $\pm$ 0.11\\
 2.75 & 0.58 $\pm$ 0.21 & 0.18 & 0.50 $\pm$ 0.23\\
\hline
 1.25 & 0.73 $\pm$ 0.04 & 0.22 & 0.06 $\pm$ 0.05& & 0.40 $\pm$ 0.09 & 0.76 & 0.11 $\pm$ 0.09 & Williams \\
 1.75 & 0.61 $\pm$ 0.06 & 0.23 & 0.23 $\pm$ 0.07\\
 2.25 & 0.60 $\pm$ 0.09 & 0.22 & 0.31 $\pm$ 0.11\\
 2.75 & 0.58 $\pm$ 0.21 & 0.18 & 0.51 $\pm$ 0.22\\
\hline
 1.25 & 0.77 $\pm$ 0.04 & 0.22 & 0.05 $\pm$ 0.04& & 0.29 $\pm$ 0.09 & 0.76 & 0.08 $\pm$ 0.08 & CATS \\
 1.75 & 0.67 $\pm$ 0.05 & 0.23 & 0.16 $\pm$ 0.08\\
 2.25 & 0.71 $\pm$ 0.09 & 0.22 & 0.30 $\pm$ 0.11\\
 2.75 & 0.67 $\pm$ 0.19 & 0.18 & 0.32 $\pm$ 0.24\\
\hline
 1.25 & 0.77 $\pm$ 0.04 & 0.22 & 0.05 $\pm$ 0.04& & 0.37 $\pm$ 0.09 & 0.76 & 0.08 $\pm$ 0.07 & Zitrin \\
 1.75 & 0.67 $\pm$ 0.06 & 0.23 & 0.25 $\pm$ 0.07\\
 2.25 & 0.71 $\pm$ 0.08 & 0.22 & 0.24 $\pm$ 0.12\\
 2.75 & 0.67 $\pm$ 0.18 & 0.18 & 0.33 $\pm$ 0.22\\
\hline
 1.25 & 0.70 $\pm$ 0.04 & 0.22 & 0.06 $\pm$ 0.05& & 0.29 $\pm$ 0.09 & 0.76 & 0.08 $\pm$ 0.08 & Sharon \\
 1.75 & 0.54 $\pm$ 0.05 & 0.23 & 0.13 $\pm$ 0.08\\
 2.25 & 0.55 $\pm$ 0.09 & 0.22 & 0.31 $\pm$ 0.11\\
 2.75 & 0.52 $\pm$ 0.15 & 0.18 & 0.17 $\pm$ 0.16\\
\hline    

	\label{table:size-mass}	
	\end{tabular}
\end{table*}

%table3
\begin{table}
	\centering
	\caption{Size evolution of galaxies at a fixed stellar mass in the form of $\propto(1+z)^{\beta_z}$ and $\propto H(z)^{\beta_h}$, as shown in Figure~\ref{img:size_evo}.}
\begin{tabular}{lcc} 
\hline
\hline
Lens model & $\beta_z$ & $\beta_h$  \\
 \hline	        
 Brada\v{c} & -1.05 $\pm$ 0.37 & -0.80 $\pm$ 0.37\\
 Williams & -0.84 $\pm$ 0.40 & -0.64 $\pm$ 0.38\\
 CATS & -1.27 $\pm$ 0.35 & -1.03 $\pm$ 0.40\\
 Zitrin & -1.17 $\pm$ 0.37 & -0.91 $\pm$ 0.39\\
 Sharon & -1.36 $\pm$ 0.28 & -1.18 $\pm$ 0.36\\
\hline
\label{table:size-evo}	
\end{tabular}
\end{table}

%figure4

%figure5

%figure6

%section5
\section{Discussion}\label{sec:discussion}

%subsection5.1
\subsection{Assessment of strong lensing uncertainties}\label{sec:lensmodel}

\begin{figure*}
\includegraphics[width=2.\columnwidth]{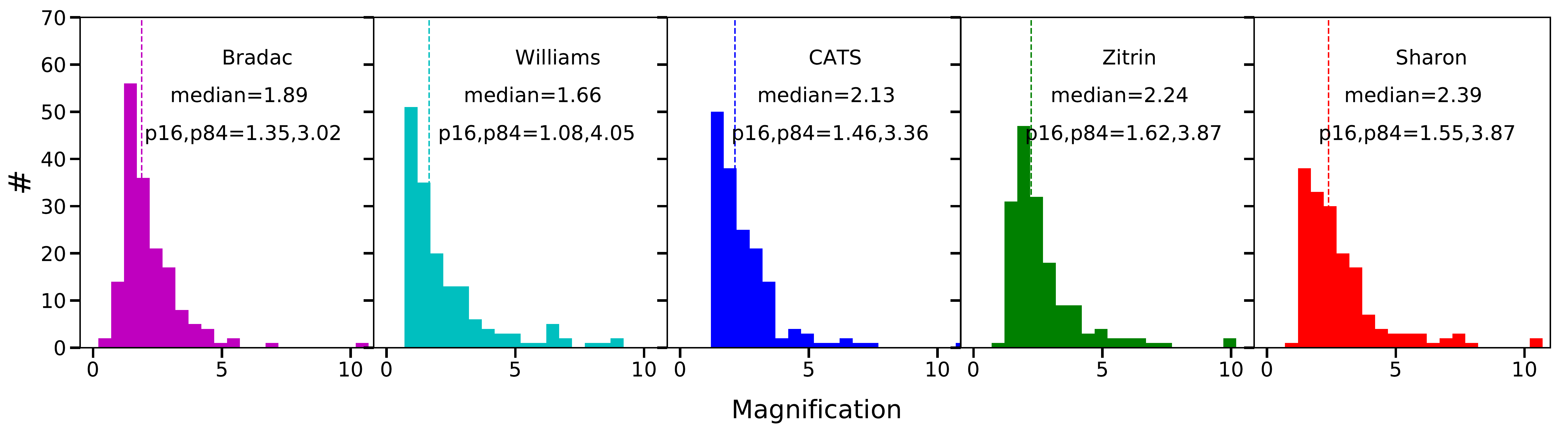}
\caption{Distribution of magnification factors derived from all five lens models. 
In each panel, we indicate the median magnification and the 16th and 84th percentiles of the distribution. }
\label{img:mag-models}
\end{figure*}

There are discrepancies among selected five lens models, although
those models are generally in agreement with each other ~\citep{Meneghetti2017}.
The choice of the lens model affects the size measurement of the intrinsic source, hence, the evolution of the size.
The magnification evaluated from five lens models differ, as presented in Figure~\ref{img:mag-models}.
We characterize the distributions of the magnification by their median and of their 16th and 84th percentiles, p16 and p84. 
The median magnification values of 4/5 models peak around $\sim$2, 
although the Williams model has a skewed distribution with smaller median value $\sim$1.66.

We show the distribution of the ratio  between size measured from Brada\v{c} model and from other lens models in Figure~\ref{img:size-hist}.
We characterize the distributions in the same manner as  Figure~\ref{img:mag-models}.
The results show that the median value of ratios are nearly around $\sim$1, but Williams model have larger median value$\sim$1.22.
The size reconstructed of same source from the Williams model is larger than from others models also shown in \cite{Yang2020}. 
The scatter in median value across models is approximately 3.7\% for 4/5 models and goes up to 11.1\% when including the Williams model.
CATS, Zitrin, and Sharon models are quite in good agreement with the Brada\v{c} model. For example,  the inter-percentile ranges (p16-p84) found among those models are
0.41, 0.50 and 0.41 respectively. 
The Williams model has more scatter than the others with an inter-percentile 0.92.

For each lens model, the results of size evolution in form of $\propto (1+z)^{\beta_z}$ and $ \propto H(z)^{\beta_h}$ are slightly different, 
see Figure~\ref{img:size_evo} and Table~\ref{table:size-evo}.
The ${\beta_z}$ and ${\beta_h}$ are $\sim$-1.2 and -1.0 across models.
Considering the large uncertainties in both ${\beta_z}$ and ${\beta_h}$, i.e., $\sim$0.35, 
we did not observe the evident difference of evolution among models.

%subsection5.2
\subsection{Comparison with previous work}

\begin{figure*}
\includegraphics[width=2.\columnwidth]{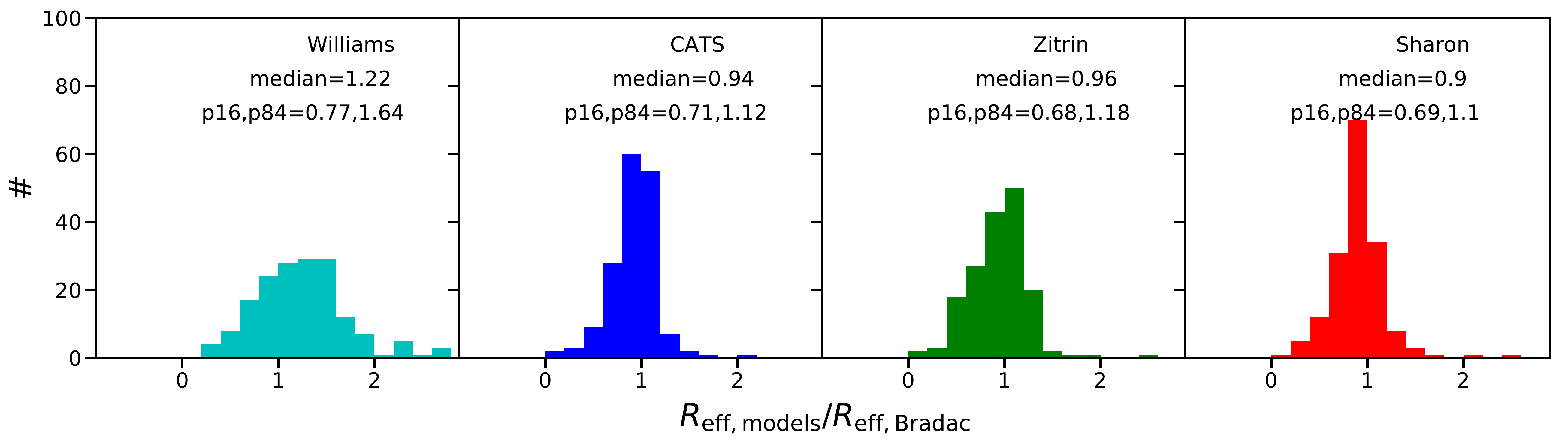}
\caption{Distribution of the ratio between the size measured using the Brada\v{c} lens model and the other lens models used in this work. In each panel, we indicate the median size and the 16th and 84th percentiles of the distribution.}
\label{img:size-hist}
\end{figure*}

Figure~\ref{img:bradac-sm} shows that our results are in excellent agreement with previous studies based on blank fields \citep{vanderwel2014,Mowla2019}. 
The agreement is important for two reasons. 
First, compared with the higher angular resolution afforded by lensing magnification, HST has allowed us to measure the size of essentially all resolved galaxies at 1$<z<$3. In this context, our work provides a test of the robustness of the results in blank fields. Second, our results show that the uncertainties associated with the lensing models do not add significant bias or scatter, as indicated in Figure~\ref{img:size-hist}. Therefore, we conclude that lensing is a valuable and well calibrated tool for the study of more compact galaxies at higher redshift.

The same agreement cannot be said, however, for results based on galaxies selected to be strongly lensed. For instance, \citet{Newton2011} and \citet{Oldham2017} both found that their sources where significantly more compact that those found in blank fields. Their sample of strongly lensed galaxies with typical magnification $\sim$10 evidently favoured more compact galaxies which are more likely to have high magnification ("magnification bias"). Future work applying a lensing selection may be able to discover higher redshift counterparts of similarly compact galaxies to those seen by \citet{Newton2011} and \citet{Oldham2017}, which are perhaps overlooked by general purpose catalogs at early selection stages when a star/galaxy separation is performed.

%subsection5.3
\subsection{Implications for the high-redshift luminosity function}

Characterizing the size distribution of galaxies is of great importance for understanding cosmic reionization \citep[e.g.][]{2012Grazian}. Galaxies are believed to be the main source of photons responsible for reionizing the intergalactic medium, provided that the luminosity function has a steep faint-end and extends to luminosities fainter than what can currently be probed by HST in blank fields. The measurement of the faint-end slope depends critically on the size distribution of galaxies, through the corrections for incompleteness approaching the detection limit. Gravitational telescopes help reach fainter intrinsic luminosities than in blank fields \citep{Kawamata2018}, provided that magnification is correctly accounted for in the cosmic volume estimates \citep[e.g.][]{Atek2018}, and the intrinsic source size can be modeled or measured simultaneously \citep[e.g.][]{Kawamata2015,Kawamata2018}. \citet{Bouwens2017} showed that the difference in the faint-end slope between smaller ($\sim$7.5mas) and larger mean size ($\sim$120mas) can be as dramatic as $\sim$0.7.

\begin{figure}
\includegraphics[width=0.8\columnwidth]{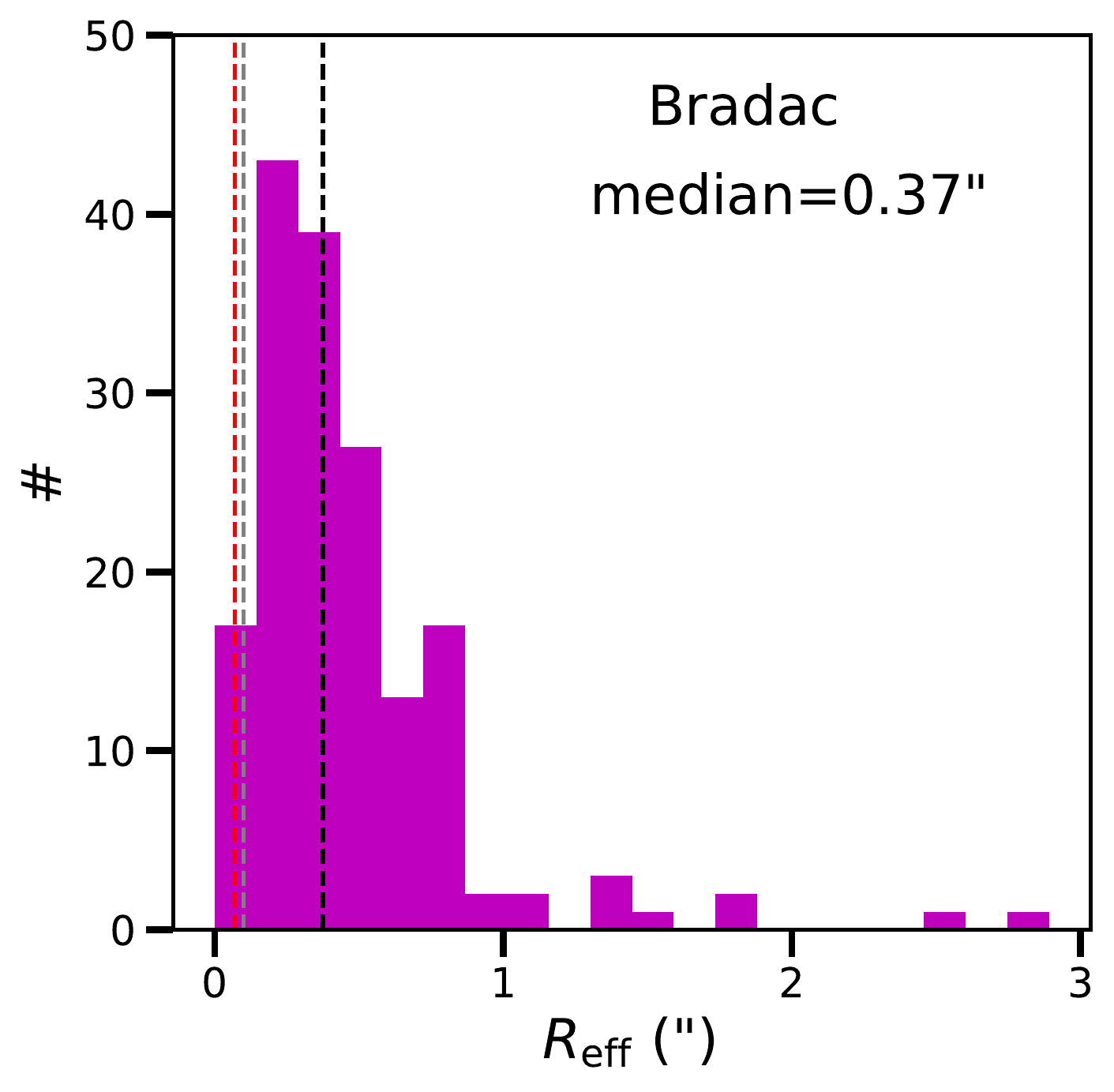}
\caption{Size distribution in units of arc-second of our sample reconstructed via the Brada\v{c} model.
The grey and red dash lines demonstrate the angular resolution of HST-WFC3/IR and JWST (0.07 arcsec at 2$\mu$m).}
\label{img:arcsize-hist}
\end{figure}

Figure~\ref{img:arcsize-hist}, shows our inferred size distribution at redshifts well below cosmic reionization. Our results are robust to the choice of initial lens model, and are the necessary stepping stone for carrying out a joint study of the galaxy size luminosity relation and luminosity function beyond the HFF and in the near future with the James Webb Space Telescope (JWST). The angular resolution of JWST (0.07 arcsec at 2$\mu$m; \citealt{Gardner2006}), coupled with gravitational telescopes should be able to pin down both the size distribution and faint-end of the luminosity function to sufficient precision to establish whether galaxies can in principle provide enough photons to reionize the universe (setting aside escape fraction uncertainty).

%figure7

%figure8

%figure9

%section6
 \section{Summary}\label{sec:summary}
 
We have studied the size-mass relation and size evolution of galaxies  
lensed by six HFF clusters 
with intrinsic (corrected for magnification) stellar mass $M_{*}>3\times10^{9}M_{\sun}$ at 1$<z<$2.5 and $M_{*}>5\times10^{9}M_{\sun}$ at 2.5$<z<$3. 
The sample is selected using multi-wavelength photometric catalogs from ASTRODEEP,
and includes spectroscopically-confirmed multiply-imaged systems.
We also evaluate the uncertainties related to the choice of lens models.
We summarize our main results as follows:

\begin{itemize}
\item We utilized the rest-frame UVJ color diagram to separate our sample into early- and late-type galaxies and build a size-mass plane for the two populations. For both populations we find results in excellent agreement with previous studies from blank fields \citep[e.g.,][]{vanderwel2014}.

\item We describe the size evolution of late-type galaxies at fixed stellar mass with the form of $(1+z)^{\beta_{z}}$ and $H(z)^{\beta_{h}}$,
 and for the fiducial lens model we find $\beta_{z}$=-1.05$\pm$0.37 and $\beta_{h}$=-0.80$\pm$0.37.
         
\item We quantify the uncertainties arising from the lensing correction by comparing 5 different publicly available models. The sizes (and size-mass relation) based on the 5 lens models agree well with each other.
Four of the models are quite in good agreement with each other: the median value of size ratio is $\sim$1, the scatter of the median is$\sim$3.7\% holistically,
and the scatter per galaxy is approximately 25\%.
One of the models produces sizes that are approximately 20\% larger than the other four. Including that model, the scatter of the median size ratio between models increases to 11\%. 
\end{itemize}

The agreement between the inferred size-mass relation with and without foreground lensing is encouraging for both endeavors. 
On the one hand, the lensing work provides a higher resolution confirmation of the blank field studies, 
suggesting any population of ultra-compact galaxies is not a significant fraction of the total. 
On the other hand, the agreement with the blank field works increases the confidence that magnification corrections are sufficiently accurate and precise for the purpose of determining the size-mass relation of galaxies.
Building on this successful comparison, in future work we plan to apply \ToolName\ , and hence \texttt{lenstronomy}, to the determination of the size-luminosity relation of galaxies at $z>7$, and its implication for the faint-end slope of the galaxy luminosity function.

\section*{Acknowledgements}
This work utilizes gravitational lensing models produced by PIs Brada\v{c}, Natarajan \& Kneib (CATS), Merten \& Zitrin, Sharon, Williams, Keeton, Bernstein and Diego, and the GLAFIC group. This lens modeling was partially funded by the HST Frontier Fields program conducted by STScI. STScI is operated by the Association of Universities for Research in Astronomy, Inc. under NASA contract NAS 5-26555. The lens models were obtained from the Mikulski Archive for Space Telescopes (MAST).  
LY is supported from the China Scholarship Council. LY and TT acknowledge support by NASA through grant JWST-ERS-1324.
The authors thank Marco Castellano, Adriano Fontana, Karl Glazebrook, Danilo Marchesini for several discussions that helped shaped the manuscript.

\section*{Data Availability} 

The data underlying this article are available in the article itself and in its online supplementary material.

%%%%%%%%%%%%%%%%%%%%%%%%%%%%%%%%%%%%%%%%%%%%%%%%%%

%%%%%%%%%%%%%%%%%%%% REFERENCES %%%%%%%%%%%%%%%%%%

% The best way to enter references is to use BibTeX:

\bibliographystyle{mnras}
\bibliography{smhff} % if your bibtex file is called example.bib

% Alternatively you could enter them by hand, like this:
% This method is tedious and prone to error if you have lots of references
%\begin{thebibliography}{99}
%bibitem[\protect\citeauthoryear{Author}{2012}]{Author2012}
%Author A.~N., 2013, Journal of Improbable Astronomy, 1, 1
%\bibitem[\protect\citeauthoryear{Others}{2013}]{Others2013}
%Others S., 2012, Journal of Interesting Stuff, 17, 198
%\end{thebibliography}

%%%%%%%%%%%%%%%%%%%%%%%%%%%%%%%%%%%%%%%%%%%%%%%%%%

%%%%%%%%%%%%%%%%% APPENDICES %%%%%%%%%%%%%%%%%%%%%

%%%%%%%%%%%%%%%%%%%%%%%%%%%%%%%%%%%%%%%%%%%%%%%%%%
%\appendix
%\section{Selected sample catalogue} 

% Don't change these lines
\bsp	% typesetting comment
\label{lastpage}
\end{document}